\documentclass[lettersize,journal]{IEEEtran}

\usepackage{amsmath,graphics,amssymb,epsfig,subfigure,color,cite,bm}
\usepackage{array}
\usepackage{multirow}
\usepackage{enumerate}
\usepackage{algorithm}
\usepackage{algpseudocode}
\usepackage{algcompatible}
\usepackage{tablefootnote}
\usepackage[flushleft]{threeparttable}

\algblockdefx{FORP}{ENDFORP}[1]%
  {\textbf{for }#1 \textbf{do in parallel}}%
  {\textbf{end}}
\algblockdefx{FOR}{ENDFOR}[1]%
  {\textbf{for }#1 \textbf{do}}%
  {\textbf{end}}
\algblockdefx{IF}{ENDIF}[1]%
  {\textbf{if }#1 \textbf{then}}%
  {\textbf{end}}
\algblockdefx{PROC}{ENDPROC}[1]
    {\textbf{procedure }#1 {}}
    {\textbf{end procedure}}
  
\algnewcommand\algorithmicprocedure{\textbf{function}}
\algnewcommand\FUNC{\item[\algorithmicprocedure]}%
\algnewcommand\algorithmicendprocedure{\textbf{end function}}
\algnewcommand\ENDFUNC{\item[\algorithmicendprocedure]}%

\usepackage{float}
\usepackage{makecell}
\usepackage[caption=false,font=normalsize,labelfont=sf,textfont=sf]{subfig}

\newcommand{\argmax}{\operatornamewithlimits{argmax}}
\newcommand{\argmin}{\operatornamewithlimits{argmin}}

\makeatletter

\newcommand{\vast}{\bBigg@{4.5}}
\newcommand{\Vast}{\bBigg@{7.5}}

\makeatother

\allowdisplaybreaks

\begin{document}
    \title{MIMO Detection under Hardware Impairments: Data Augmentation With Boosting}

	\author{Yujin Kang, Seunghyun Jeon, Junyong Shin, Yo-Seb Jeon, \IEEEmembership{Member,~IEEE}, \\ and H. Vincent Poor, \IEEEmembership{Life Fellow,~IEEE}
		\thanks{Yujin Kang, Seunghyun Jeon, Junyong Shin, and Yo-Seb Jeon are with the Department of Electrical Engineering, POSTECH, Pohang, Gyeongbuk 37673, Republic of Korea (e-mail: yujinkang @postech.ac.kr; seunghyeon.jeon@postech.ac.kr; sjyong@postech.ac.kr; yoseb.jeon@postech.ac.kr).}
        \thanks{H. Vincent Poor is with the Department of Electrical Engineering, Princeton University, Princeton, NJ 08544 (e-mail: poor@princeton.edu).}
	}
	\vspace{-2mm}	
	
	\maketitle
	\vspace{-12mm}

	\begin{abstract}
         This paper addresses a data detection problem for multiple-input multiple-output (MIMO) communication systems with hardware impairments. To facilitate maximum likelihood (ML) data detection without knowledge of nonlinear and unknown hardware impairments, we develop novel likelihood function (LF) estimation methods based on data augmentation and boosting. The core idea of our methods is to generate multiple augmented datasets by injecting noise with various distributions into seed data consisting of online received signals. We then estimate the LF using each augmented dataset based on either the expectation maximization (EM) algorithm or the kernel density estimation (KDE) method. Inspired by boosting, we further refine the estimated LF by linearly combining the multiple LF estimates obtained from the augmented datasets. To determine the weights for this linear combination, we develop three methods that take different approaches to measure the reliability of the estimated LFs. Simulation results demonstrate that both the EM- and KDE-based LF estimation methods offer significant performance gains over existing LF estimation methods. Our results also show that the effectiveness of the proposed methods improves as the size of the augmented data increases.
	\end{abstract}
	

    \section{Introduction}\label{Sec:Intro}
    Driven by the explosive growth in demand for higher data rates, the wireless communication industry has been propelled towards exploring higher frequency bands, particularly the millimeter-wave (mmWave) and terahertz (THz) bands. This shift offers vast bandwidth and the potential for unprecedented data rates \cite{mmwave_key, mmwave}. However, it also introduces new challenges in system design. One of the most significant challenges is the increased impact of hardware impairments on system performance. As we move to mmWave and THz bands, the use of ideal RF components becomes impractical due to the cost and hardware limitations. A representative example is nonlinear distortion in power amplifiers (PAs), which occurs when the input signal exceeds the amplifier's linear operating range. This effect becomes particularly severe in high-frequency bands, such as mmWave and THz \cite{rf_distortion, rf_imperfection}. Moreover, conventional high-resolution ADCs exhibit a linear increase in power consumption as signal bandwidth and the number of received RF chains grow \cite{adc_survey}. To circumvent the power consumption problem, drastically reducing ADC resolution emerges as a solution for power savings. Unfortunately, the impact of these non-ideal hardware components poses a substantial challenge to the operation of conventional communication systems, particularly in multiple-input multiple-output (MIMO) systems, which are crucial for achieving high spectral efficiency \cite{capacity_limits_HI}. 

    Acknowledging the inevitability of hardware impairments, data detection methods for communication systems with hardware impairments have been studied in the literature \cite{thz_HI, further_HI, jinsung, optdet_HI, cognitive_IQ, iq_imbalance_detector}. For instance, data detection methods in  \cite{optdet_HI, cognitive_IQ, iq_imbalance_detector} focused on enhancing the detection performance based on prior knowledge regarding the effects of hardware impairments. By characterizing precise impairment models, these methods lead to significant improvements in data detection performance even under the effects of hardware impairments. However, it is crucial to acknowledge that in real-world scenarios, obtaining such perfect knowledge is challenging due to the non-generalizable hardware behavior and the intricate interactions of various impairment sources. Consequently, while these studies offer important theoretical foundations, there remains a compelling demand for realizing detection methods that do not rely on prior knowledge. 
    To overcome the limitations due to lack of knowledge on hardware impairments, machine-learning approaches have been considered in recent studies \cite{lowcomplexity_HI, dl_OTFS_HI, dithering_onebit, admm_HI, magazine_AI}. 
    These approaches aim to learn the effects of unknown hardware impairments directly from data, without requiring accurate knowledge about these impairments. Specifically, deep neural networks, trained on large datasets of impaired signals, can learn detection strategies that outperform traditional methods in various scenarios \cite{onebit_deep, deep_qpsk}. Despite the benefits of machine learning in terms of performance and universality, the machine-learning-based methods often falter when confronted with real-world wireless scenarios in which the environments presumed during training differ from online communication environments.
    

    To tackle these limitations and improve performance in dynamic and unpredictable scenarios, online-learning-based approaches have been proposed in \cite{supervised_ADC, semisup_detection_gan, jinman}. The common idea of these approaches is to make the data detection methods adaptable to online communication environments, such as channel variations and hardware impairments, based on online received signals. For instance, the method in \cite{supervised_ADC} utilized repeated transmissions of data signals to generate sufficient training samples for the system. However, this method requires a prohibitively large signaling overhead that may not be feasible in practical systems. 
    To address this problem, online received signals from unknown data have been utilized as {\em online} samples for training learning-based detection methods \cite{semisup_detection_gan, jinman}. Specifically, in \cite{semisup_detection_gan}, pilot signals, which have the same form as the transmitted signals, were utilized as training data. However, in practical systems, the design of pilot signals can differ from modulated symbols used for data transmission, making it difficult to support diverse transmission schemes. This limitation was tackled in \cite{jinman} which does not require extra pilot overhead or specific designs for the pilot signals. Although this method offers impressive adaptability to the unknown effects of hardware impairments, it suffers from performance degradation due to limited dataset size and overfitting issues. 

    Data augmentation has emerged as a promising solution to address the data shortages and overfitting problems by artificially expanding a training dataset through the creation of new, synthetic samples \cite{augmentation_survey}. One of the primary advantages of data augmentation is its ability to improve the generalization capabilities of models. Introducing a variety of transformations to the training dataset, such as rotations and scaling, data augmentation enables models to learn from a broader range of scenarios. This diversity helps mitigate overfitting, ensuring that the model does not become overly specialized to the training data alone \cite{augmentation_magazine, ae_augmentation, augmentation_precoding, modulation_HI, data_augmentation_deep}.
    Data augmentation for communications with hardware impairments has been studied in the context of modulation \cite{modulation_HI}, demonstrating its potential to enhance the robustness of algorithms against hardware-induced distortions. Additionally, the impact of data augmentation on deep neural network-based receivers was investigated in \cite{data_augmentation_deep}. However, these methods require substantial pilot data transmission to generate seed information for augmentation, which serves as the true labeled dataset for each block, presenting a limitation in practical applications. 
    Despite the complementary advantages of online data generation and data augmentation, none of the existing studies has explored the combination of these two techniques for communications involving hardware impairments.

     In this paper, we present a novel data detection approach for MIMO communication systems with hardware impairments that leverages the complementary advantages of both online data generation and data augmentation. In our approach, we utilize a partial set of online received signals as {\em online} seed data and generate augmented data samples through noise injection. These augmented samples are utilized to estimate the likelihood function (LF) of the system with hardware impairments. Specifically, we create multiple augmented datasets using various noise distributions and apply boosting to combine multiple LF estimates, effectively addressing both data shortages and overfitting issues. Through this approach, we offer a practical solution for estimating the LFs of the system under unknown hardware impairments, without requiring additional pilot signals or prior knowledge. Simulation results demonstrate that the maximum likelihood (ML) data detection using our LF estimates outperforms existing approaches in the presence of hardware impairments.
    The key contributions of this paper can be outlined as follows:
    
    \begin{itemize}
        \item 
        We propose a novel data augmentation framework for LF estimation in time-invariant MIMO systems with hardware impairments. A key feature of our framework is its use of online received signals as {\em online} seed data for data augmentation, eliminating the need for extra pilot signals or prior knowledge. Additionally, our framework generates multiple augmented datasets using various noise distributions with different parameters and then leverages boosting to make effective use of these multiple datasets. Incorporating data augmentation with boosting has not been previously considered in the literature. Through this unique approach, our framework addresses both data shortage and overfitting problems in the LF estimation for MIMO systems with hardware impairments. We demonstrate that our framework can be successfully combined with two LF estimation methods: (i) the EM algorithm \cite{em_algorithm}, and (ii) the KDE method in \cite{kde}. 

        
        
        \item 
        We extend the proposed data augmentation framework to LF estimation in time-varying MIMO systems with hardware impairments. The core idea behind this extension is to segment a communication block into multiple sub-blocks and then apply the proposed framework sequentially to each sub-block. Specifically, the LF estimates from the previous sub-block are used to initialize the LF estimation in the current sub-block. Our extended framework realizes a concept of successive online learning, for the first time, in time-varying MIMO systems with hardware impairments, requiring no additional pilot signals beyond those needed for initial channel estimation. This simple yet practical approach successfully broadens the applicability of our framework to more complex and realistic scenarios.

        \item
        Using simulations, we demonstrate the superiority of the proposed data augmentation framework in MIMO systems with non-ideal PAs and low-resolution ADCs. Our results show that ML data detection using our framework outperforms ML detection with existing LF estimation methods, particularly at high signal-to-noise ratio (SNR) regimes. Moreover, this performance gain increases as the size of augmented dataset increases, highlighting the efficacy of data augmentation combined with boosting. Simulation results also demonstrate that our framework designed for time-varying systems offers a significant performance improvement over existing methods in time-varying scenarios.
    \end{itemize}


    \vspace{2mm}
    {\em Notation:} Upper-case and lower-case boldface letters denote matrices and column vectors, respectively. 
    $\mathbb{E[\cdot]}$ is the statistical expectation, $(\cdot)^{\sf T}$ is the transpose, and $(\cdot)^{\sf H}$ is the conjugate transpose. $|\mathcal{A}|$ is the cardinality of set $\mathcal{A}$. 
    $\boldsymbol{a}_i$ represents the $i$-th element of vector $\boldsymbol{a}$. 
    $\|\boldsymbol{a}\| = \sqrt{\boldsymbol{a}^{\sf T}\boldsymbol{a}}$ is the Euclidean norm of a real vector $\boldsymbol{a}$. 
    $\|\boldsymbol{a}\|_{1} = \Sigma_i | \boldsymbol{a}_{i}|$ is the L1-norm of a real vector $\boldsymbol{a}$.
    $\mathcal{CN}({\bm \mu},{\bf R})$ represents the distribution of a circularly symmetric complex Gaussian random vector with mean vector ${\bm \mu}$ and covariance matrix ${\bf R}$. 
    $\boldsymbol{0}_n$ is an $n$-dimensional vector whose elements are zero and ${\bf I}_N$ is an $N$ by $N$ identity matrix.
    ${\rm Re}(\cdot)$ and ${\rm Im}(\cdot)$ are the real and imaginary parts of a complex value, respectively.

    \section{System Model}
\begin{figure}
        \centering
        {\epsfig{file=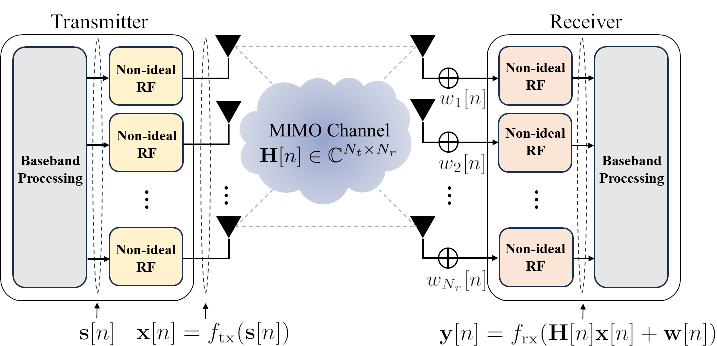,width=8.5cm}}
        \caption{An illustration of a MIMO communication system under hardware impairments.} 
        \label{fig:system_model}\vspace{-3mm}
    \end{figure}

    Consider a MIMO communication system figured in Fig. \ref{fig:system_model} in which a transmitter equipped with $N_t$ antennas transmits information data to a receiver equipped with $N_r$ antennas. 
   
    \subsection{Channel Model}
    Assume that the MIMO channel between the transmitter and receiver is modeled as frequency-flat Rayleigh fading channels with unit variance.  
    Let ${\bf H}[n]\in\mathbb{C}^{N_r\times N_t}$ be a MIMO channel matrix between the transmitter and receiver at time slot $n\in\mathcal{T}\triangleq\{1,2,\ldots,T\}$, where $T$ is the total duration of a communication block.
    We consider two types of channel models: (i) time-invariant channels, and (ii) time-varying channels. For the time-invariant channels, we assume that the entries of the MIMO channel matrix ${\bf H}[n]$ remain constant during $T$ time slots, i.e., ${\bf H}[n]={\bf H}$, $\forall n\in\mathcal{T}$. 
    For the time-varying channels, we assume that the MIMO channel matrix changes over time according to the first-order Gauss-Markov process: 
    \begin{align}
        {\bf H}[n] = \zeta {\bf H}[n-1] + \sqrt{1-\zeta^2}{\bf V}[n], 
    \end{align}
    where the entries of ${\bf V}[n]$ are independent and follow a circularly symmetric complex Gaussian distribution with zero mean and unit variance.

    \subsection{Signal Model with Hardware Impairments}
    The first $T_{\rm p}$ time slots of the communication block are dedicated to transmitting pilot signals which are known at the receiver for the purpose of channel estimation. The remaining $T_{\rm d} = T- T_{\rm p}$ time slots are allocated for data transmission. Let ${\bf s}_{\rm p}[n] = \big[s_{{\rm p},1}[n],\cdots,s_{{\rm p},N_t}[n]\big]^{\sf T}\in\mathbb{C}^{N_t}$ be a pilot signal vector transmitted at time slot $n\in\mathcal{T}_{\rm p}\triangleq\{1,\ldots, T_{\rm p}\}$. Assume that each entry of the pilot signal vector ${\bf s}_{\rm p}[n]$ has a unit power (i.e., $\mathbb{E}[| s_{{\rm p},i}[n]|^2]=1$). Also, let ${\bf s}[n] = \big[s_1[n],\cdots,s_{N_t}[n]\big]^{\sf T}\in\mathcal{X}^{N_t}$ be a data symbol vector transmitted at time slot $n\in\mathcal{T}_{\rm d}\triangleq\{T_{\rm p}+1,\ldots,T\}$, where $\mathcal{X}$ is a constellation set $\mathcal{X}$ such that $\mathbb{E}[| s_i[n]|^2]=1$.

    In this work, we assume that both the transmitter and the receiver are equipped with {\em non-ideal} RF components. 
    Let $f_{\rm tx}:\mathbb{C}^{N_t}\rightarrow \mathbb{C}^{N_t}$ be a transmit (TX) distortion function which represents the combined effect of the non-ideal RF components at the transmitter.
    Then, a transmitted signal at time slot $n$ is expressed as 
    \begin{align}
        {\bf x}[n] = 
        \begin{cases}
            f_{\rm tx}({\bf s}_{\rm p}[n]), & n\in\mathcal{T}_{\rm p}, \\
            f_{\rm tx}({\bf s}[n]), & n\in\mathcal{T}_{\rm d}.
        \end{cases}
    \end{align}
    Similarly, let $f_{\rm rx}:\mathbb{C}^{N_r}\rightarrow \mathbb{C}^{N_r}$ be a receive (RX) distortion function which represents the combined effect of the non-ideal RF components at the receiver.
    Then, a distorted received signal at time slot $n$ is given by 
    \begin{align}\label{eq:received_signal}
        {\bf y}[n] &= f_{\rm rx}\big({\bf H}[n]{\bf x}[n]+{\bf w}[n]\big) \\
        &=\begin{cases}
            f_{\rm rx}\big({\bf H}[n]f_{\rm tx}({\bf s}_{\rm p}[n])+{\bf w}[n]\big), & n\in\mathcal{T}_{\rm p}, \\
            f_{\rm rx}\big({\bf H}[n]f_{\rm tx}({\bf s}[n])+{\bf w}[n]\big), & n\in\mathcal{T}_{\rm d},
        \end{cases}
    \end{align}
    where ${\bf w}[n]\in\mathbb{C}^{N_r}$ is additive white Gaussian noise (AWGN) distributed as ${\bf w}[n] \sim \mathcal{CN}({\bf 0}_{N_r},\sigma^2 {\bf I}_{N_r})$.

    \subsection{TX and RX Distortion Functions}
    To provide a more realistic illustration of the TX and RX distortion functions, we present examples of these functions in a scenario where non-ideal PAs are employed at the transmitter and low-resolution ADCs are employed at the receiver. Note that this scenario has been considered in much of the existing literature (e.g., \cite{jinman}, \cite{elm}). In this scenario, the TX distortion function captures the effect of the non-ideal PAs and is therefore given by
    \begin{align}\label{eq:PA}
        f_{\rm tx}({\bf s}[n]) = \big[{\sf PA}({\bf s}_1[n]), {\sf PA}({\bf s}_2[n]), \cdots, {\sf PA}({\bf s}_{N_{\rm t}}[n]) \big]^{\sf T},    
    \end{align}
    for $n\in\mathcal{T}_{\rm d}$, where ${\sf PA}(x) = A(\vert x\vert)e^{j\{\angle(x) + \Phi(\vert x\vert)\}}$, $\angle(x)$ is the angle of $x$, 
    \begin{align}
        A(\vert x\vert) &= \frac{\alpha_a\vert x\vert}{1+\epsilon_a\vert x\vert^2},~~~
        \Phi(\vert x\vert) = \frac{\alpha_\phi \vert x\vert^2}{1+\epsilon_\phi\vert x\vert^2},
    \end{align}
    and $\alpha_a$, $\epsilon_a$, $\alpha_\phi$, $\epsilon_\phi$ are PA-dependent parameters \cite{pa_model}.
    Meanwhile, the RX distortion function captures the effect of the low-resolution ADCs and is therefore expressed as 
    \begin{align}\label{eq:ADC}
        f_{\rm rx}({\bf r}[n]) = &\big[{\sf ADC}({\sf Re}\{{\bf r}_1[n]\}), \cdots, {\sf ADC}({\sf Re}\{{\bf r}_{N_{\rm r}}[n]\}) \big]^{\sf T} \nonumber \\ 
        &+ j
        \big[{\sf ADC}({\sf Im}\{{\bf r}_1[n]\}), \cdots, {\sf ADC}({\sf Im}\{{\bf r}_{N_{\rm r}}[n]\}) \big]^{\sf T},
    \end{align} 
    where $r_k={\sf ADC}(r)$ if $b_{k-1} < r \leq b_k$, $r_k$ is the $k$-th quantization output of the ADC, and $b_k$ is the $k$-th quantization boundary of the ADC.

    \subsection{Likelihood Function (LF) of the Considered System}\label{Sec:LF}
    The statistical behavior of a distortion signal introduced by the TX and RX distortion functions can be characterized as follows:
    The distorted received signal in \eqref{eq:received_signal} for $n\in \mathcal{T}_{\rm d}$ can be rewritten as 
    \begin{align}
        {\bf y}[n] 
        &= {\bf H}[n]{\bf s}[n] + f_{\rm rx}\big({\bf H}[n]f_{\rm tx}({\bf s}[n])+{\bf w}[n]\big) - {\bf H}[n]{\bf s}[n] \nonumber \\
        &= {\bf H}[n]{\bf s}[n] + {\bf d}[n],
    \end{align}     
    where ${\bf d}[n]$ is the distortion signal at time slot $n \in \mathcal{T}_{\rm d}$ defined as 
    \begin{align}\label{eq:distortion_signal}
        {\bf d}[n] = f_{\rm rx}\big({\bf H}[n]f_{\rm tx}({\bf s}[n])+{\bf w}[n]\big) - {\bf H}[n]{\bf s}[n].
    \end{align}     
    The above expression shows that the distortion signal is a function of the MIMO channel matrix ${\bf H}[n]$, the data symbol vector ${\bf s}[n]$, and the AWGN signal ${\bf w}[n]$.
    Recall that the data symbol vector is a member of a finite set $\mathcal{X}^{N_{\rm t}}$. 
    Let ${\bf s}_k$ be the $k$-th element of the set $\mathcal{X}^{N_{\rm t}}$ for $k\in \mathcal{K}\triangleq \{1,\ldots,K\}$, where $K\triangleq |\mathcal{X}|^{N_{\rm t}}$.
    Then the data symbol vector ${\bf s}[n]$ satisfies ${\bf s}[n] = {\bf s}_{k}$ for a certain $k\in \mathcal{K}$. 
    Utilizing this fact, for the time-varying channels, the LF of the distorted received signal given that ${\bf s}[n]={\bf s}_k$ is expressed as
    \begin{align}\label{eq:LF_def}
        p_{k,n}\big({\bf y}[n]\big) &\triangleq p\big({\bf y}[n] \big| {\bf s}[n] = {\bf s}_k\big)  \nonumber \\
        &=p\big({\bf H}[n]{\bf s}_k + {\bf d}[n] \big| {\bf s}[n] = {\bf s}_k \big) \nonumber \\
        &=p\big(f_{\rm rx}\big({\bf H}[n]f_{\rm tx}({\bf s}_k)+{\bf w}[n]\big)\big).
    \end{align}
    For the time-invariant channels, the MIMO channel matrix ${\bf H}[n]$ can be treated as a constant matrix, implying that the randomness of the distortion signal stems solely from the AWGN signal ${\bf w}[n]$.
    Since the distribution of ${\bf w}[n]$ does not change over time, for the time-invariant channels, the LF in \eqref{eq:LF_def} satisfies that $p_{k,n}\big({\bf y}[n]\big) = p_{k}\big({\bf y}[n]\big)$ for $n\in\mathcal{T}$. 
    Once the LFs are computed, the optimal ML data detection for minimizing the detection error probability is expressed as 
    \begin{align}\label{eq:MLD}
        \hat{\bf s}_{\rm ML}[n] =\underset{{\bf s}_k \in \mathcal{X}^{N_t}}{\argmax} ~p({\bf y}[n]|{\bf s}_k).
    \end{align}

    As can be seen in the above expression, to maximize the detection performance at the receiver, it must have perfect knowledge of the LFs, denoted by $\{p_k({\bf y}[n])\}_{k=1}^K$.
    Unfortunately, acquiring this knowledge is highly challenging, not only because of the nonlinearities in the distortion functions but also due to the lack of information about these functions. Therefore, to devise the effective data detection for MIMO systems with hardware impairments, it is crucial to enable accurate estimation of the LFs even in the presence of nonlinear and unknown distortion functions.

    \section{Basic LF Estimation Methods}\label{Sec:Basic}
    Before presenting the proposed LF estimation methods, in this section, we first introduce two basic LF estimation methods for MIMO systems with hardware impairments.
    These methods will serve as the foundation for the development of the proposed approaches. In this section, we assume that the receiver has access to a sample dataset, denoted by $\mathcal{D}_{\bf y} = \{{\bf y}^{(1)},{\bf y}^{(2)},\ldots,{\bf y}^{(D)}\}$, consisting of distorted received signal samples from the system. How to generate this dataset will be discussed in Sec. IV.

    \subsection{EM-based LF Estimation}\label{Sec:EM}
    The first method, referred to as an {\em EM-based} method, approximates the LF $p_{k}\big({\bf y}[n]\big)$ using a multivariate complex Gaussian distribution with a symbol-specific mean vector ${\bm \mu}_k$ and covariance matrix ${\bm \Sigma}_k$, i.e., 
    \begin{align}\label{eq:p_k_d_EM}
        &p_k({\bf y}[n]) \approx 
        \mathcal{CN}\big({\bf y}[n];{\bm \mu}_k,{\bm \Sigma}_k\big)  \nonumber \\
        &\triangleq \frac{1}{\pi^{N_{\rm rx}}|{\bm \Sigma}_k|} \exp\left(-({\bf y}[n]-{\bm \mu}_k)^{\sf H}{\bm \Sigma}_k^{-1}({\bf y}[n]-{\bm \mu}_k)\right).
    \end{align}
    Given ${\bf s}[n]={\bf s}_k$, the entries of ${\bf y}[n]$ are statistically uncorrelated because the entries of ${\bf w}[n]$ are uncorrelated. 
    Considering this fact, we further assume that the covariance matrix ${\bm \Sigma}_k$ is a diagonal matrix with different diagonal entries, i.e., 
    \begin{align}\label{eq:p_k_d_EM}
        {\bm \Sigma}_k = {\sf diag}(\sigma_{k,1}^2, \sigma_{k,2}^2,\ldots, \sigma_{k,N_{ r}}^2).
    \end{align}
    It should be noted that this model is a generalization of the existing models in \cite{jinman}.
    The aforementioned modeling strategy provides a tractable form for the LF of the MIMO systems with hardware impairments. However, the LF in \eqref{eq:p_k_d_EM} is still unknown at the receiver due to the lack of information about a parameter ${\bm \theta} = ({\bm \theta}_1,{\bm \theta}_2,\ldots, {\bm \theta}_K)$, where ${\bm \theta}_{k} = ({\bm \mu}_k, {\bm \Sigma}_k)$.

    To estimate the unknown parameter ${\bm \theta}$, we leverage a sample dataset $\mathcal{D}_{\bf y}$ based on the EM algorithm \cite{em_algorithm}. 
    The EM algorithm is an iterative algorithm that estimates unknown parameters from observed data, when the observed data is associated with hidden random variables.
    To leverage this algorithm, we introduce an indicator random variable $z_{n,k}$ defined as $z_{n,k} = \mathbb{I}[{\bf s}[n] = {\bf s}_k]$, where $\mathbb{I}[A]$ is an indicator function that outputs $1$ if the event $A$ is true and $0$ otherwise. 
    The random variable $z_{n,k}$ is treated as an unobserved hidden data because this random variable is unknown at the receiver. 
    Now, we define the complete data, which consists of both the observed and hidden data, as $({\bf Y},{\bf Z})$, where ${\bf Y} = ({\bf y}^{(1)},{\bf y}^{(2)},\ldots, {\bf y}^{(D)})$ and ${\bf Z}$ is an $D\times K$ matrix whose $(n,k)$-th entry is given by $z_{n,k}$. 
    Then the likelihood of the complete data for the parameter ${\bm \theta}$ is given by \cite{prml}
    \begin{align}
        p\big({\bf Y},{\bf Z} | {\bm \theta}\big) 
        &= \prod_{n=1}^{D} \prod_{k=1}^K \pi_{n,k}^{z_{n,k}} \mathcal{CN}\big({\bf y}[n];{\bm \mu}_k,{\bm \Sigma}_k\big)^{z_{n,k}},
    \end{align}
    where $\pi_{n,k} \triangleq \mathbb{P}[z_{n,k}=1] = \mathbb{P}[{\bf s}[n]={\bf s}_k]$. 

    The EM algorithm iterates between the E-step and M-step to update the parameter ${\bm \theta}$ based on the observed data ${\bf Y}$. Let ${\bm \theta}^{(i)} = ({\bm \theta}_1^{(i)},{\bm \theta}_2^{(i)},\ldots, {\bm \theta}_K^{(i)})$ be the parameter estimated at iteration $i$, where ${\bm \theta}_{k}^{(i)} = ({\bm \mu}_k^{(i)}, {\bm \Sigma}_k^{(i)})$. In the E-step, the expectation of the log-likelihood of the complete data is computed for given observed data ${\bf Y}$ and the current parameter ${\bm \theta}^{(i)}$. The result is given by 
    \begin{align}
        &Q({\bm \theta} | {\bm \theta}^{(i)}) 
        = \mathbb{E}\big[\ln p({\bf Y},{\bf Z}|{\bm \theta}) \big| {\bf Y}, {\bm \theta}^{(i)} \big] \nonumber \\
        &= \sum_{n=1}^{D} \sum_{k=1}^K \hat{z}_{n,k}^{(i)} \big\{ \ln \pi_{n,k} + \ln \mathcal{CN}\big({\bf y}^{(n)}; {\bm \mu}_k,{\bm \Sigma}_k\big) \big\},
    \end{align}
    where 
    \begin{align}\label{eq:em_estep}
        \hat{z}_{n,k}^{(i)}  
        &= \mathbb{E}\big[ z_{n,k} \big| {\bf y}^{(n)}, {\bm \theta}^{(i)} \big] \nonumber \\
        &= \frac{\pi_{n,k}^{(i)} \mathcal{CN}\big({\bf y}^{(n)};  {\bm \mu}_k,{\bm \Sigma}_k\big) }
        {\sum_{j=1}^K \pi_{n,j}^{(i)} \mathcal{CN}\big({\bf y}^{(n)}; {\bm \mu}_j,{\bm \Sigma}_j\big)}.
    \end{align}    
    In the M-step, the parameter that maximizes $Q({\bm \theta} | {\bm \theta}^{(i)})$ is computed and then set as a new parameter for iteration $i+1$. The result is given by \cite{prml} 
    \begin{align}\label{eq:em_mstep}
        \pi_{k}^{(i+1)} &= \frac{1}{D}\sum_{n=1}^D \hat{z}_{n,k}^{(i)}, \\
        {\bm \mu}_{k}^{(i+1)} &= \frac{1}{\sum_{n=1}^D \hat{z}_{n,k}^{(i)}}
        \sum_{n=1}^D \hat{z}_{n,k}^{(i)}{\bf y}^{(n)},\\
        {\bm \Sigma}_{k}^{(i+1)} &= \frac{1}{\sum_{n=1}^D \hat{z}_{n,k}^{(i)}} \nonumber \\
        &\times {\sf diag} \left(\sum_{n=1}^D \hat{z}_{n,k}^{(i)} \big({\bf y}^{(n)} -{\bm \mu}_{k}^{(i+1)}  \big) \big( {\bf y}^{(n)} - {\bm \mu}_{k}^{(i+1)} \big)^{\sf H}
        \right).
    \end{align}
    The EM algorithm iterates until a specific stopping criterion is met or until it reaches a predefined number of iterations.


    \subsection{KDE-based LF Estimation}\label{Sec:KDE}

    A limitation of the EM-based LF estimation method in Sec.~\ref{Sec:EM} is that it heavily relies on a generalized Gaussian model to characterize the effects of hardware impairments on the distorted received signal. In practical systems, some of these effects can only be modeled using non-Gaussian distributions due to severe nonlinearities and/or time-varying characteristics. In such cases, the EM-based method fundamentally suffers from model mismatch, leading to a degradation in LF estimation accuracy.

    To circumvent the limitation of the EM-based method, the second method leverages kernel density estimation (KDE) \cite{kde}, which provides a framework for approximating the PDF of a random variable directly from training samples. Unlike the EM-based method, the KDE method does not rely on specific probabilistic distributions to model the LFs. Rather, this method utilizes training samples to approximate the LF at a given point ${\bf y}[n]$ by integrating the contributions of all samples to that point.
    This is achieved by summing kernel functions centered on each observed data point, resulting in a smooth approximation of the underlying PDF.

    Suppose that we have a set of received signal samples labeled with the $k$-th data symbol vector ${\bf s}_k$, denoted by $\mathcal{D}^{k}_{\bf y}$. 
    Then, the KDE function to approximate the LF associated with the $k$-th data symbol vector ${\bf s}_k$ is defined as \cite{kde}
   \begin{align}\label{eq:KDE}
         &{\sf KDE}_k({\bf y}[n];\mathcal{D}^{k}_{\bf y}) \triangleq \frac{1}{|\mathcal{D}^{k}_{\bf y}|} \sum_{{\bf y}^\prime \in \mathcal{D}^{k}_{\bf y}} {\phi}_{\bf M} \left( {\bf y}^{\prime} - {\bf y}[n]\right) \nonumber \\ 
         &~~~~~~~~~~~~~~~~= \frac{|{\bf M}|^{-1}}{|\mathcal{D}^{k}_{\bf y}|} \sum_{{\bf y}^\prime \in \mathcal{D}^{k}_{\bf y}}  {\phi}\big( {\bf M}^{-1/2} ({\bf y}^{\prime}  - {\bf y}[n])\big), 
    \end{align}
    where ${\phi}(\cdot)$ is a kernel function,
    ${\bf M}$ is a bandwidth matrix controlling the estimated density's smoothness and $| {\bf M}|$ is the determinant of matrix ${\bf M}$.
    Utilizing the KDE functions, we approximate the LF $p_{k}\big({\bf y}[n]\big)$ as follows:
    \begin{align}\label{eq:p_k_d_KDE}
        &p_k({\bf y}[n]) \approx {\sf KDE}_k({\bf y}[n];\mathcal{D}^{k}_{\bf y}),~~k\in\mathcal{K}.
    \end{align}

    To achieve reliable density estimation using the KDE method, it is essential to select an appropriate kernel function ${\phi}(\cdot)$.
    The most widely adopted kernel is a Gaussian kernel which provides a smooth estimate of density and ensures robustness against Gaussian noise. For a multivariate case, the complex Gaussian kernel is defined as
    \begin{align}\label{eq:kernel}
      {\phi}({\bf x}) \triangleq \frac{1}{\pi^{N_r}}\exp\left( - \lVert{\bf x}\rVert^2 \right).
    \end{align}
    Motivated by its robustness and generality, we employ the multivariate complex Gaussian kernel to estimate the LF $p_{k}\big({\bf y}[n]\big)$, which yields
    \begin{align}\label{eq:Gaussian_KDE}
        & {\sf KDE}_k({\bf y}[n];\mathcal{D}^{k}_{\bf y}) =\frac{|{\bf M}|^{-1}}{|\mathcal{D}^{k}_{\bf y}|\pi^{N_r}} \nonumber \\
        &~~~\times  \sum_{{\bf y}^\prime \in \mathcal{D}^{k}_{\bf y}}  \exp\left(-({\bf y}^\prime - {\bf y}[n])^{\sf H} {\bf M}^{-1} ({\bf y}^\prime - {\bf y}[n])\right).
    \end{align}
    As can be seen in \eqref{eq:p_k_d_KDE} and \eqref{eq:Gaussian_KDE}, the KDE-based LF estimation does not make specific assumptions about the LFs. Rather, this method mimics the behavior of the true LFs by utilizing data samples that inherently capture the nonlinear effects of hardware impairments.

    


    \subsection{Challenges Behind the LF Estimation}\label{Sec:LF_Challenge}
    To employ the LF estimation approaches in Sec. III-A and Sec. III-B, the receiver needs to have a sufficient number of received signal samples associated with each symbol vector. A straightforward approach to achieve this goal is to directly transmit a known training sequence consisting of repeated symbol vectors at the transmitter, as suggested in \cite{supervised_ADC}. However, this approach requires a large communication overhead, which not only increases communication latency but also degrades spectral efficiency. 
    An alternative approach developed in \cite{jinman} is to utilize {\em online} received signals, $\{{\bf y}[n]\}_{n\in\mathcal{T}_{\rm d}}$, associated with transmitted data symbol vectors as a sample dataset, where the outputs of a symbol detector are regarded as labels for these signals.
    However, in practical systems, the number of available online received signals within a channel coherence time is limited. 
    Additionally, utilizing online received signals with incorrect labels can lead to overfitting, as the sample dataset is identical to the testing dataset in this approach \cite{jinman}.
    To facilitate accurate LF estimation in practical communication systems, it is critical to devise an effective solution to address both the data shortage and overfitting problems. 

    \section{The Proposed LF Estimation Methods based on \\ Data Augmentation with Boosting}\label{Sec:Pro_TI}
    In this section, we propose novel LF estimation methods for MIMO systems with hardware impairments, which leverage both data augmentation and boosting to address the challenges of LF estimation discussed in Sec.~\ref{Sec:LF_Challenge}.

    \subsection{Overview of the Proposed Methods}
\begin{figure*}
        \centering
        {\epsfig{file=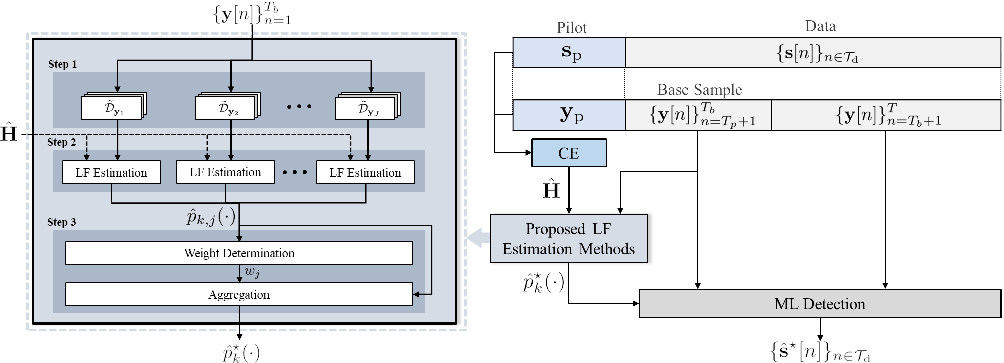,width=15cm}}
        \caption{Block diagram of the data detection process with the proposed LF estimation method for time-invariant channels.} 
        \label{fig:inv_block}\vspace{-3mm}
    \end{figure*}

    The proposed LF estimation methods encompass three key steps: (i) data augmentation, (ii) parallel LF estimation, and (iii) boosting. 
    The high-level procedure of these steps is illustrated in Fig. \ref{fig:inv_block} and described below.
     
    \begin{enumerate}[(i)]
    \item {\bf Step 1 (Data Augmentation):} This step aims to generate sample dataset for estimating the LFs of the system. To this end, we first designate a portion of the online received signals as {\em base} samples which serve as the foundation (seed data) for data augmentation. We then augment the base samples through noise injection. In particular, we generate $J$ augmented datasets by employing $J$ distinct noise distributions. 
     
    \item {\bf Step 2 (Parallel LF Estimation):} This step aims to estimate the LFs of the system based on the generated datasets. To this end, we utilize each augmented dataset to estimate the LFs of the system based on the EM method (Sec.~\ref{Sec:EM}) or the KDE method (Sec.~\ref{Sec:KDE}). As a result, we obtain $J$ different sets of LF estimates because the augmented datasets have distinct sample distributions.   
    
    \item {\bf Step 3 (Boosting):} This step aims to compute the {\em refined} estimates for the LFs of the system. We set the refined LF estimate as a weighted combination of $J$ LF estimates obtained in Step 2. To properly combine these estimates, we devise three methods to determine the weights based on the reliability of the LF estimates.
    \end{enumerate}



    \subsection{Step 1: Data Augmentation}\label{Sec:Pro_DA}
    The goal of the data augmentation step is to generate a proper sample dataset $\mathcal{D}_{\bf y}$ that can be utilized for LF estimation. To avoid the need for extra training overhead, we focus on utilizing online received signals $\{{\bf y}[n]\}_{n\in\mathcal{T}_{\rm d}}$ associated with transmitted data symbol vectors. Unfortunately, the direct use of these signals as a sample dataset leads to data shortage and overfitting problems, as discussed in Sec. III-C. To address these issues, we develop a data augmentation strategy to increase the number of received signal samples while avoiding overfitting.

    We start by generating {\em base} samples which serve as the foundation (seed data) for data augmentation. 
    Let $T_{\rm b} \leq T_{\rm d}$ be the number of the base samples. 
    In our strategy, we designate the first $T_{\rm b}$ received signals as the base samples because this choice minimizes the latency required for generating the base samples. Then the set of time indices belonging to the base sample is expressed as $\mathcal{T}_{\rm b}\triangleq\{T_{\rm p}+1,\dots,T_{\rm p}+T_{\rm b}\}$.
    After generating the base samples, we generate $I_{\rm DA}$ augmented samples for each base sample by artificially injecting noise signals which are randomly drawn from a certain noise distribution. The $i$-th augmented sample for the base sample ${\bf y}[n]$ is expressed as  
    \begin{align}\label{eq:aug}
          {\bf y}^{(d)} &= {\bf y}[n] + \mathbf{n}^{(i)},   \:  n\in \mathcal{T}_{\rm b}, i \in \mathcal{I}_{\rm DA},
    \end{align}
    where the index of augmented sample is $d \triangleq (n-T_p-1)I_{DA}+i$, the index set is defined as $\mathcal{I}_{\rm DA}\triangleq \{1,\ldots,I_{\rm DA}\}$, and ${{\mathbf{n}}^{(i)}}$ is the $i$-th random noise signal drawn from a certain noise distribution for the base sample ${\bf y}[n]$.


    The primary goal of our data augmentation is to generate a resilient and diverse sample dataset for received signals to enable accurate LF estimation in the presence of unknown and unpredictable hardware impairments. To achieve this, the probability distribution for generating artificial noise signals must be carefully chosen. However, optimizing this distribution is intractable without prior knowledge of the hardware impairments (i.e., TX and RX distortion functions). To circumvent this challenge, we employ multiple noise distributions to generate augmented samples, each with a different distribution, allowing us to potentially capture diverse effects of the hardware impairments. We specifically select the following three distributions for noise sampling: (i) Gaussian, (ii) uniform, and (iii) Laplace distributions, motivated by their widespread application in modeling various phenomena in communication systems. Details of each distribution are provided below.

    \begin{itemize}
        \item {\bf Gaussian distribution:} This is the circularly symmetric complex Gaussian distribution, which is commonly used to model thermal noise signals. The PDF of this distribution is given by 
    \begin{align}\label{eq:gaussian_noise}
        f({\mathbf{n}}) = \frac{1}{(\pi\sigma_{\rm G}^2)^{N_r}}\exp\bigg( - \frac{\lVert {\mathbf{n}} \rVert^2}{\sigma_{\rm G}^2}\bigg),
    \end{align}
    where $\sigma_{{\rm G}}^2$ is the variance of the Gaussian distribution, and ${\mathbf{n}} \sim \mathcal{CN}({\bf 0},{\sigma_{{\rm G}}^2 {\bf I}})$. 

        \item {\bf  Uniform distribution:} This is the uniform distribution, which is useful for enforcing arbitrary behavior in the injected noise signals. Let ${n}_r$ be the $r$-th entry of the noise signal ${\bf n}$. The PDF of this distribution is given by 
    \begin{align}\label{eq:uniform_noise}
        f({\mathbf{n}}) =  
        \begin{cases}
        1/{\sigma_U^{2N_r}}, &\!\! {\rm if}~ {\rm Re}({n}_r), {\rm Im}({n}_r) \in [-\frac{\sigma_U}{2}, \frac{\sigma_U}{2}], \forall r, \\
        0, &\!\! {\rm otherwise},
        \end{cases}
    \end{align}
    where $\sigma_U$ determines the boundaries for the real and imaginary parts of each entry of the noise signal. 

    \item {\bf Laplace distribution:} This is the Laplace distribution, which becomes a reasonable choice when impulsive noise is prevalent because this distribution has a sharper peak compared to the Gaussian distribution. The PDF of this distribution is given by 
    \begin{align}\label{eq:laplace_noise}
        f({\mathbf{n}}) =  \frac{1}{(2\sigma_L)^{2N_r}}\exp\bigg( - \frac{\lVert {\rm Re}({\mathbf{n}})\rVert_1 + \lVert{\rm Im}({\mathbf{n}}) \rVert_1}{\sigma_L}\bigg),
    \end{align}
    where $\sigma_L$ represents the scale parameter of the Laplace distribution.
    \end{itemize}

    Note that the Gaussian, uniform, and Laplace distributions are fully characterized by the parameters $\sigma_{G}$, $\sigma_{U}$, and $\sigma_{L}$, respectively. Since different parameters yield different noise distributions, we predetermine $J/3$ distinct parameters for each of the three distributions, where $J$ is the total number of noise distributions considered in our data augmentation strategy. As a result, we obtain $J$ distinct noise distributions. Each noise distribution is used to generate a distinct augmented dataset, as explained above. The $i$-th augmented sample generated using the $j$-th noise distribution is given by  
    \begin{align}\label{eq:augmentation}
        \{{\bf y}^{(d)}_j\}_{d=1}^{I_{DA}T_{\rm b}} = \{ {\bf y}[n] + \mathbf{n}^{(i)}_{j} |  \: n\in \mathcal{T}_{\rm b}, i \in \mathcal{I}_{\rm DA} \} 
    \end{align}
    for $j \in \{1,\cdots,J\}$, where ${\bf n}^{(i)}_j$ is the $i$-th random noise signal drawn from the $j$-th noise distribution for the base sample ${\bf y}[n]$. Consequently, the $j$-th augmented sample dataset is defined as 
    $\mathcal{D}_{{\bf y}_j} = \{ {\bf y}^{(d)}_j | d\in\{1,2,\cdots,I_{\rm DA}T_{\rm b}\} \}$.
    Through this process, we yield $J$ distinct sample datasets that can be utilized for the LF estimation in the subsequent step. 

    \subsection{Step 2: Parallel LF Estimation}\label{Sec:Pro_LF}
    The next step of the proposed methods is to estimate the LFs of the system by utilizing $J$ augmented datasets, ${\mathcal{D}}_{{\bf y}_1},\ldots,{\mathcal{D}}_{{\bf y}_J}$, obtained in the previous step. 
    In particular, we utilize these $J$ datasets in parallel, where each dataset is used independently to determine its own LF estimates for the same system.
    For the LF estimation, we consider the EM and KDE methods, introduced in Sec.~\ref{Sec:EM} and Sec.~\ref{Sec:KDE}, respectively. 
    In the following, we illustrate how the augmented dataset is utilized by each of these two approaches. 

    \begin{itemize}
    \item {\bf EM-based LF estimation:}
        In this approach, we leverage the EM algorithm from Sec.~\ref{Sec:EM} for LF estimation. To initialize the EM algorithm, we use a channel estimate, denoted as $\hat{\bf H}$, obtained from a conventional pilot-aided channel estimation method that ignores the effects of hardware impairments. The EM parameter for the $k$-th symbol vector is initialized as ${\bm \theta}_{k}^{(0)} = ({\bm \mu}_k^{(0)}, {\bm \Sigma}_k^{(0)}) = (\hat{\bf H} {\bf s}_k, \sigma^2{\bf I})$, where ${\bm \mu}_k^{(0)}$ and ${\bm \Sigma}_k^{(0)}$ correspond to the mean vector and covariance matrix of the received signal when transmitting the $k$-th symbol vector. Note that the mean vector $\hat{\bf H} {\bf s}_k$ and the covariance matrix $\sigma^2{\bf I}$ are exact when there are no hardware impairments (i.e., $f_{\rm rx}(x)=f_{\rm tx}(x) =x$). After initialization, we perform the EM algorithm in parallel using each of the $J$ augmented datasets. Let $I_{\rm EM}$ be the number of iterations for the EM algorithm. Then the LF estimate for the $k$-th symbol vector using the $j$-th augmented dataset ${\mathcal{D}}_{{\bf y}_j}$ is given by
        \begin{align}\label{eq:em_lf}
            \hat{p}_{k,j}({\bf y}[n]) =
            \mathcal{CN}\big({\bf y}[n];{\bm \mu}_{k,j}^{(I_{\rm EM})},{\bm \Sigma}_{k,j}^{(I_{\rm EM})}\big),
        \end{align}
        for $j\in\{1,\ldots,J\}$, where ${\bm \theta}_{k,j}^{(I_{\rm EM})} = ({\bm \mu}_{k,j}^{(I_{\rm EM})},{\bm \Sigma}_{k,j}^{(I_{\rm EM})})$ is the parameter determined by the EM algorithm using the $j$-th augmented dataset ${\mathcal{ D}}_{{\bf y}_j}$ after $I_{\rm EM}$ iterations.

    \item {\bf KDE-based LF estimation:}
        Unlike the EM-based LF estimation, implementation of the KDE-based LF estimation in Sec.~\ref{Sec:KDE} necessitates labeled received signal samples. In our methods, however, the label of the received signal sample corresponds to the index of the transmitted symbol vector which is unknown to the receiver. To circumvent this difficulty, we coarsely label the augmented samples based on a conventional ML data detection method that ignores the effects of hardware impairments. Then the noisy label of the augmented sample ${\bf y}^{(d)}_j$ is determined as
        \begin{align}\label{eq:coarse_MLD}
             \check{k}^{(d)}_j &= \underset{{k\in \{1,\ldots,K\}}}{{\argmax}}~\mathcal{CN}\big({\bf y}^{(d)}_j ;\hat{\bf H}{\bf s}_k, \sigma^2{\bf I} \big) \nonumber \\
            &= \underset{{k\in \{1,\ldots,K\}}}{{\argmin}} \lVert {\bf y}^{(d)}_j -\hat{\bf H}{\bf s}_k\rVert^2,
        \end{align}
        where  $\hat{\bf H}$ is a channel estimate determined by a conventional pilot-aided channel estimation method that ignores the effects of hardware impairments. 
        
        ~~~Now, we construct the set from the augmented samples and their noisy labels as 
        $\check{\mathcal{D}}_{{\bf y}_j} = \{(\{ {\bf y}^{(d)}_j, \check{k}^{(d)}_j \}) \}$. 
        This labeled dataset serves as an input for the KDE-based LF estimation. To approximate the PDF of each ${p}_{k,j}({\bf y}[n])$, we partition this set based on identical labels. Let $\check{\mathcal{D}}_{{\bf y}_j}^k$ be the subset of $\check{\mathcal{D}}_{{\bf y}_j}$, which consists of the augmented samples that are labeled as $k$ according to \eqref{eq:coarse_MLD}, i.e.,
        \begin{align}
            \check{\mathcal{D}}_{{\bf y}_j}^k = \big\{   {\bf y}  \big| \check{k} = k, ({\bf y},\check{k})\in \check{\mathcal{D}}_{{\bf y}^j} \big\}.
        \end{align}
        By the definition, $\check{\mathcal{D}}_{{\bf y}_j}^1, \ldots, \check{\mathcal{D}}_{{\bf y}_j}^K$ are mutually exclusive subsets of $\check{\mathcal{D}}_{{\bf y}_j}$.
        Then the LF estimate for the $k$-th symbol vector using the $j$-th augmented dataset  $\check{\mathcal{D}}_{{\bf y}_j}$ is given by
        \begin{align}\label{eq:kde_lf}
            \check{p}_{k,j}({\bf y}[n])  = {\sf KDE}_k({\bf y}[n];\check{\mathcal{D}}_{{\bf y}_j}^{k}),
        \end{align}
        for $j\in\{1,\ldots,J\}$. 
        
        ~~~When the effect of hardware impairment is severe, the quality of the noisy labels $\{\check{k}^{(d)}_j\}$ can be poor as these labels are determined based on mismatched LFs. This may degrade the performance of the KDE-based LF estimation. To address this problem, we refine the noisy labels by utilizing the base samples as virtual pilot signals, as done in data-aided channel estimation \cite{yoseb:twc}. In this strategy, we label the base samples $\{{\bf y}[n]\}_{n\in\mathcal{T}_{\rm b}}$ based on the LF estimates in \eqref{eq:kde_lf}, i.e.,  $\check{\bf s}[n] = {\bf s}_{\check{k}_{j}[n]}$, where
         \begin{align}\label{eq:KDE_refine}
            \check{k}_{j}[n] &= \underset{{k\in \{1,\ldots,K\}}}{{\argmax}}~\check{p}_{k,j}({\bf y}[n]),~~\forall n\in\mathcal{T}_{\rm b}.
        \end{align}
        We then utilize virtual pilot signals $\{\check{\bf s}[n]\}_{n\in\mathcal{T}_{\rm b}}$ to perform a refined channel estimation. Let $\hat{\bf H}_j^\prime$ be a data-aided channel estimate obtained from the aforementioned process.
        This allows us to re-label the augmented samples by using the data-aided channel estimate as follows:
        \begin{align}\label{eq:KDE_refine_label}
             \hat{k}^{(d)}_j = \underset{{k\in \{1,\ldots,K\}}}{{\argmin}} \lVert {\bf y}^{(d)}_j -\hat{\bf H}_j^\prime {\bf s}_k\rVert^2,
        \end{align}
        After this process, the aforementioned KDE-based LF estimation is performed again based on the dataset $\hat{\mathcal{D}}_{{\bf y}_j} = \{(\{ {\bf y}^{(d)}_j, \hat{k}^{(d)}_j ) \}$ with the refined labels.

        \end{itemize}

       Our parallel LF estimation step produces $J$ estimates, given by $\hat{p}_{k,1}({\bf y}[n]),\ldots, \hat{p}_{k,J}({\bf y}[n])$, to approximate the unknown true LF $p_{k}({\bf y}[n])$. These estimates differ across the augmented datasets due to the distinct sample distributions. This fact implies that the quality of the LF estimates may also differ among the datasets. Consequently, an additional step is required to effectively utilize these LF estimates based on their quality, which will be discussed in the following subsection.

    \subsection{Step 3: Boosting}\label{Sec:Pro_Boost}
    The final step of the proposed methods is to compute the {\em refined} estimates for the LFs of the system by utilizing the $J$ LF estimates obtained in the previous step. 
    Our key idea is to set the refined estimate as a weighted combination of the $J$ estimates determined using the $J$ augmented datasets.
    Based on this idea, the refined LF estimate, namely $\hat{p}_k^\star({\bf y}[n])$, is expressed as 
    \begin{align}\label{eq:ensemble_algorithm}
        \hat{p}_k^\star({\bf y}[n]) = \sum^{J}_{j=1} w_{j} \hat{p}_{k,j}({\bf y}[n]),
    \end{align}
    where $w_j$ is the weight assigned to the estimate determined using the $j$-th augmented dataset.

    To maximize the accuracy of the refined LF estimate, it is crucial to assign appropriate weights based on the quality of the LF estimates. Unlike conventional methods, such as \cite{bem}, which enhance performance by forming an ensemble with known true targets, our approach addresses scenarios where the true target (i.e., the true LF) is unknown. This requires the development of a new strategy to evaluate the quality of the estimates without relying on ground truth information. Therefore, we propose the following weight determination methods that do not require any prior knowledge about the system.
    \begin{itemize}
        \item {\bf Uniform Aggregation:}
        The simplest, yet practical, method is to assign equal weights to all the estimates by assuming that they have the same quality. 
        In this case, the weight $w_j$ is expressed as
        \begin{align}\label{eq:aggregation_uniform}
            w_j = \frac{1}{J}.
        \end{align}
        The advantage of this strategy is that it minimizes the computational complexity for determining the weights as it does not require any extra process.  

        \item {\bf Probabilistic Aggregation:}
        The data detection can be considered as $K$-way classification in which the output of data detection can only take one of $K$ possible outcomes. In this context, $\hat{k}$ is modeled as a categorical random variable with $K$ possible categories. Based on this perspective, the probabilistic aggregation method evaluates the quality of the LF estimate using the probability of the detection outputs. To achieve this, the detection outputs for the base samples are computed by utilizing each set of LF estimates. Specifically, for $j\in\{1,\ldots,J\}$, the ML data detection is performed by substituting the true LFs with the $j$-th LF estimates, $\{\hat{p}_{k,j}({\bf y})\}_k$. Then the detection output for the base sample ${\bf y}[n]$ with $n\in \mathcal{T}_{\rm b}$ is given by 
        \begin{align}
            \hat{k}_j[n] = \underset{{k\in \{1,\ldots,K\}}}{{\argmax}}~ \hat{p}_{k,j}({\bf y}[n]),~ n\in \mathcal{T}_{\rm b}.
        \end{align} 
        Consequently, the ratio of the detection output $k$ among $T_{\rm b}$ outputs in $\{\hat{k}_j[n] \}_{n\in\mathcal{T}_{\rm b}}$ is given by 
        \begin{align}\label{eq:r_jk}
            r_{k,j} = \frac{c_{k,j}}{\sum_{l=1}^{K} {c}_{l,j}},~~k\in\mathcal{K},~j\in\{1,\ldots,J\}, 
        \end{align}
        where $c_{k,j}$ is the number of the base samples detected as $k$.
        The Dirichlet distribution models the probability distribution for a $K$-way categorical event. Utilizing this distribution, the $j$-th weight is determined to be proportional to the PDF for a $K$-dimensional probability vector ${\bf r}_j=[r_{1,j}\cdots,r_{K,j}]^{\sf T}$. 
        As a result, the weight $w_j$ is determined as
        \begin{align}\label{eq:aggregation_dirichlet}
            w_j = \frac{w_j^\prime}{\sum_{l=1}^J w_l^\prime},~~\text{where}~~w_j^\prime = \frac{1}{B(\boldsymbol{\alpha})} \prod_{k=1}^K r_{k,j}^{\alpha_{k} -1},
        \end{align}
         $\boldsymbol{\alpha} \triangleq \{\alpha_{1},\cdots,\alpha_{K}\}$ is a concentration parameter, and $B(\boldsymbol{\alpha})$ is a normalizing constant.
        
        \item {\bf Max Aggregation:}
        The key idea of this method is to choose the best LF estimate that produces the most probable distribution for the detection outputs. Similar to our probabilistic aggregation strategy, the detection outputs for the base samples are computed from the ML data detection which substitutes the true LFs with each set of the LF estimates. Let $\{\hat{k}_j[n] \}_{n\in\mathcal{T}_{\rm b}}$ be the set of the detection outputs for the base samples when using the $j$-th LF estimates, $\{\hat{p}_{k,j}({\bf y})\}_k$, for $j\in\{1,\ldots, J\}$. Then the index of the LF estimates associated with the most probable detection outputs is determined as 
        \begin{align}\label{eq:aggregation_Jstar}
            {J}^\star = \argmax_j \frac{1}{B(\boldsymbol{\alpha})} \prod_{k=1}^K r_{k,j}^{\alpha_{k} -1},
        \end{align}
        where $r_{k,j}$ is given in \eqref{eq:r_jk}. 
        Then, the weight $w_j$ is determined as 
       \begin{align}\label{eq:aggregation_hard}
            w_{j} = 
            \begin{cases}
                1, & j = {J}^\star\\
                0, & j \neq {J}^\star.
            \end{cases}
        \end{align}        
    \end{itemize}

    After refining the LF estimates based on one of three weight-determination methods, we perform the ML data detection in \eqref{eq:MLD} by substituting the true LFs with the refined LF estimates. The detection output for the received signal ${\bf y}[n]$ at time slot $n\in\mathcal{T}_{\rm d}$ is given by $\hat{\bf s}[n] = {\bf s}_{\hat{k}^\star[n]}$, where  
    \begin{align}\label{eq:final_ml}
        \hat{k}^\star[n] =\underset{k \in \{1,\ldots,K\}}{\argmax}~ \hat{p}_k^\star ({\bf y}[n])= \underset{{\bf s}_k \in \mathcal{X}^{N_t}}{\argmax}~\sum^{J}_{j=1} w_{j} \hat{p}_{k,j}({\bf y}[n]).
    \end{align}
    The overall detection process with the proposed LF estimation method is summarized in \textbf{Algorithm 1}.
    

    \begin{algorithm}[ht]
    \caption{Data Detection with the Proposed LF Estimation Method}\label{alg:LF Estimation}
    {\small
        {\begin{algorithmic}[1]
            \REQUIRE
            Base samples $\{{\bf y}[n]\}^{T_b}_{n=1}$, additive noise parameters $\{\sigma_G\}, \{\sigma_U\}, \{\sigma_U\}$, an estimated channel $\hat{\bf H}$, a noise power $\sigma^2$
            \ENSURE Detected symbol vectors $\hat{\bf s}[n] = {\bf s}_{\hat{k}^{\star}[n]}, n\in\mathcal{T}_{\rm d}$   

            \FOR{$j=1$ to $J$} 
                \STATE Generate an augmented dataset ${\mathcal{D}}_{{\bf y}_j}$ from \eqref{eq:augmentation}.
                \IF{EM-based LF estimation}
                    \STATE Set $({\bm \mu}_k^{(0)}, {\bm \Sigma}_k^{(0)}) = (\hat{\bf H} {\bf s}_k, \sigma^2{\bf I}), \forall k$.
                    \FOR {$i=1$ to $I_{\rm EM}$}
                        \STATE Compute $\hat{z}_{n,k}^{(i)}$ from \eqref{eq:em_estep}, $\forall n,k$.
                        \STATE Compute ${\bm \mu}_k^{(i)}$ and ${\bm \Sigma}_k^{(i)}$ from \eqref{eq:em_mstep}, $\forall k$.
                    \ENDFOR
                    \STATE Set ${\bm \theta}_{k,j}^{(I_{EM})} = ({\bm \mu}_{k,j}^{(I_{\rm EM})},{\bm \Sigma}_{k,j}^{(I_{\rm EM})}), \forall k$.
                    ~~~\STATE Compute the LF $\hat{p}_{k,j}({\bf y}[n])$ from \eqref{eq:em_lf}$, n\in\mathcal{T}_{\rm b}$.
                \ELSIF{KDE-based LF estimation}
                    \STATE Construct a labeled dataset $\check{\mathcal{D}}_{{\bf y}_j}$ from \eqref{eq:coarse_MLD}.  
                    \STATE Compute the LF $\check{p}_{k,j}({\bf y}[n])$ from \eqref{eq:kde_lf}, $n\in\mathcal{T}_{\rm b}$. 
                    \STATE Label the base samples from \eqref{eq:KDE_refine}. 
                    \STATE Calculate a refined channel estimate $\hat{\bf H}^{'}_j$ with \eqref{eq:final_ml}. 
                    \STATE Set a labeled dataset $\hat{\mathcal{D}}_{{\bf y}_j}^k,\forall k$. 
                    \STATE Compute the LF $\hat{p}_{k,j}({\bf y}[n])$ from \eqref{eq:kde_lf}, $n\in\mathcal{T}_{\rm b}$.
                \ENDIF
            \ENDFOR
            \STATE Calculate $w_j,\forall j$ from \eqref{eq:aggregation_uniform}, \eqref{eq:aggregation_dirichlet}, or \eqref{eq:aggregation_hard}.
            \STATE $\hat{p}_k^\star({\bf y}[n]) = \sum^{J}_{j=1} w_{j} \hat{p}_{k,j}({\bf y}[n]), \forall k,n$.
            \STATE Determine $\hat{k}^{\star}[n]$ from \eqref{eq:final_ml},  $n\in\mathcal{T}_{\rm d}$.
            \STATE Set $\hat{\bf s}[n] = {\bf s}_{\hat{k}^{\star}[n]}, n\in\mathcal{T}_{\rm d}$.
            
        \end{algorithmic}}} 
    \end{algorithm}


    \section{Extension of the Proposed LF Estimation Methods in Time-Varying Channels}\label{Sec:Pro_TV}
    Although the proposed LF estimation methods in Sec.~\ref{Sec:Pro_TI} effectively address both data shortage and overfitting problems, their efficacy is guaranteed only when the channels are time-invariant. In time-varying channels, however, the LFs change within the communication block, meaning the LFs estimated using the first $T_{\rm b}$ received signals differ from those for the remainder of the block. To address this limitation, in this section, we extend the proposed methods in Sec.~\ref{Sec:Pro_TI} to apply them to time-varying channels.

    \subsection{Overview of the Modified LF Estimation Process}
\begin{figure*}
        \centering
        {\epsfig{file=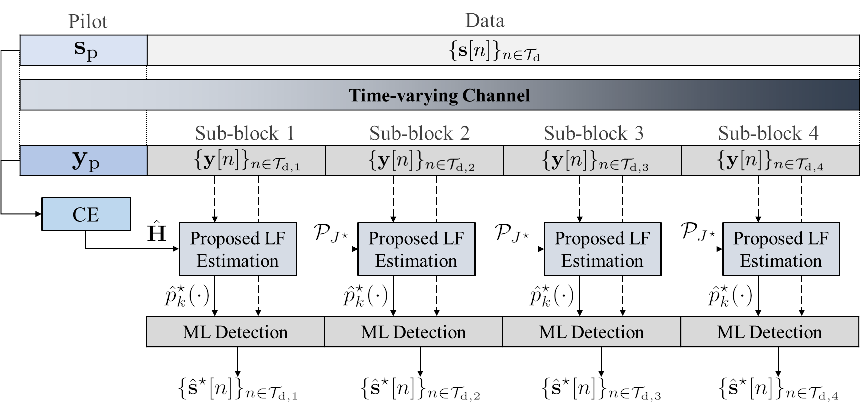, width = 11cm}}
        \caption{Block diagram of the data detection process with the proposed LF estimation method for time-varying channels.}  \vspace{-1.5mm}
        \label{fig:var_block}
    \end{figure*}

    The basic idea behind our modification is to continually update the estimated LFs across multiple sub-blocks, while applying the same methods from Sec.~\ref{Sec:Pro_TI} within each sub-block. To this end, we divide the data transmission block (consisting of $T_{\rm d}$ time slots) into $S$ sub-blocks, with each sub-block consisting of $T_{{\rm d}}^\prime = T_{\rm d}/S$ slots. We assume that the channels within each sub-block do not change significantly. Based on this assumption, we estimate the LFs by utilizing the proposed methods in Sec.~\ref{Sec:Pro_TI} within each sub-block. After determining the LF estimates in the $s$-th sub-block, these estimates are used for data detection within the current sub-block. These estimates are then passed to the $(s+1)$-th sub-block, where they are utilized as the initialization for LF estimation. In this way, we continually update the LF estimates so that they can track the temporal changes in LF values caused by temporal channel variations. The modified LF estimation process with the aforementioned continual estimation strategy is illustrated in Fig.\ref{fig:var_block}. In what follows, we describe the detailed changes in each step of the modified LF estimation process. 
    
    \subsection{Sub-Block Data Augmentation} 
    In the modified LF estimation methods, the data augmentation step is performed in a per-sub-block manner. Unlike the data augmentation step in Sec.~\ref{Sec:Pro_DA} which utilizes only the first $T_{\rm b}$ received signals as the base samples, a key modification in this step is to utilize the received signals within the entire sub-block as {\em base} samples, implying that $T_{\rm b} = T_{\rm d}^\prime$. This allows us to mitigate the data insufficiency problem in the time-varying channels. 
    The base sample, comprised of the entire sub-block, undergoes augmentation through noise injection using different noise distributions (e.g., Gaussian, uniform, and Laplace distributions) as described in Sec.~\ref{Sec:Pro_DA}. 
    To be more specific, let $\mathcal{T}_{{\rm d},s}$ be the set of time slot indices associated with the $s$-th sub-block. 
    Then the $i$-th augmentation sample for the base sample $\{{\bf y}[n]\}_{n\in\mathcal{T}_{{\rm d},s}}$ using the $j$-th noise distribution is expressed as
    \begin{align}
        {\bf y}^{(d)}_j = {\bf y}[n] + {\bf n}_{j}^{(i)}, n\in\mathcal{T}_{{\rm d},s}, i\in\mathcal{I}_{\rm DA}
    \end{align}
    for $j\in\{1,\ldots,J\}$. Consequently, the $j$-th augmented sample dataset for the $s$-th sub-block is defined as ${\mathcal{D}}_{{\bf y}_{j}} = \{ {\bf y}_j^{(d)} | d \in\{1,2,\cdots,I_{\rm DA}T_{\rm b}\}\}$.  Through the sub-block data augmentation process, we effectively obtain $J$ augmented datasets that can be utilized for the LF estimation within each sub-block.

    \subsection{Sub-Block LF Estimation}
    The next step in the modified methods is to estimate the LFs within each sub-block using the $J$ augmented datasets for the corresponding sub-block. A key modification in this step is the initialization of the LF estimation method. For time-invariant channels, we rely on a conventional pilot-aided channel estimation method to obtain the initial information for each LF estimation method, as described in Sec.~\ref{Sec:Pro_LF}. However, in time-varying channels, if the current sub-block is temporally distant from the pilot transmission block, the initialization based on pilot-aided channel estimates may suffer from outdated channel information. Such inaccurate initialization can degrade the performance of the LF estimation methods, as their effectiveness depends on the accuracy of the initial parameters (in the EM-based method) or initial labels (in the KDE-based method). To mitigate this issue, we propose transferring the most up-to-date LF information from the current sub-block to the next sub-block. The specific initialization procedure for these sub-blocks is detailed below.



    \begin{itemize}
    \item{\bf EM-based LF estimation:}
    For the initialization of the EM-based LF estimation method, we use the EM parameters determined from the previous sub-block. Specifically, for the initialization at the $(s+1)$-th sub-block, we select the best EM parameters obtained at the $s$-th sub-block. To do this, we determine the best index $J^{\star}$ among the $J$ augmented datasets using the max aggregation method given in \eqref{eq:aggregation_Jstar}. Then, the final EM parameter ${\bm \theta}_{k,J^{\star}}^{(I_{\rm EM})} = ({\bm \mu}_{k,J^{\star}}^{(I_{\rm EM})},{\bm \Sigma}_{k,J^{\star}}^{(I_{\rm EM})})$, determined from the $J^{\star}$-th augmented dataset in the $s$-th sub-block, is then used as the initial EM parameter for the $(s+1)$-th sub-block. After initialization, we perform the EM-based LF estimation in parallel using each of the $J$ augmented datasets in the same manner described in Sec.~\ref{Sec:Pro_LF}.

    
    \item{\bf KDE-based LF estimation:}
    For the initialization of the KDE-based LF estimation method, we use the labeled dataset determined from the previous sub-block. Specifically, for the initialization at the $(s+1)$-th sub-block, we select the best index $J^{\star}$ for the $s$-th sub-block among the $J$ augmented datasets by using the max aggregation method given in \eqref{eq:aggregation_Jstar}. Then, the selected dataset  is expressed as, 
    \begin{align}\label{eq:kde_Jstar}
        \hat{\mathcal{D}}_{{\bf y}_{J^{\star}}} = \{(\{ {\bf y}^{(d)}_{J^{\star}}, \hat{k}^{(d)}_{J^{\star}} ) \},
    \end{align}
    which is determined by the KDE-based LF estimation method using the  $J^{\star}$-th augmented dataset. Utilizing the dataset $\hat{\mathcal{D}}_{{\bf y}_{J^{\star}}}$ transferred from the $s$-th sub-block, the KDE-based LF estimation is performed in parallel using each of the $J$ augmented datasets in the same manner described in Sec.~\ref{Sec:Pro_LF}. 
    \end{itemize}

    \subsection{Sub-Block Boosting}
    The boosting step in the modified LF estimation method is essentially the same as the one described in Sec.~\ref{Sec:Pro_Boost}. Among three different weight determination methods, we continue using the max aggregation method, as it identifies the optimal LF estimates among the $J$ LF estimates. These optimal estimates are then used for the initialization of the LF estimation methods in the subsequent sub-block. The overall detection process with the modified LF estimation method for time-varying channels is summarized in \textbf{Algorithm 2}.
    

    \begin{algorithm}[ht]
    \caption{Data Detection with the Modified LF Estimation Method}\label{alg:LF Estimation}
    {\small
        {\begin{algorithmic}[1]
            \REQUIRE
            Received signals $\{{\bf y}[n]\}^{T_d}_{n=1}$, additive noise parameters $\{\sigma_G\}, \{\sigma_U\}, \{\sigma_U\}$, an estimated channel $\hat{\bf H}$, a noise power $\sigma^2$
            \ENSURE Detected symbol vectors $\hat{\bf s}[n] = {\bf s}_{\hat{k}^{\star}[n]}, n\in\mathcal{T}_{\rm d}$   
            \FOR{$s=1$ to $S$}
                \FOR{$j=1$ to $J$} 
                    \STATE Generate augmented sample datasets ${\mathcal{D}}_{{\bf y}_j}$ from \eqref{eq:augmentation}
                    \IF{EM-based LF estimation}
                        \IF{$s=1$}
                            \STATE Set $({\bm \mu}_k^{(0)}, {\bm \Sigma}_k^{(0)}) = (\hat{\bf H} {\bf s}_k, \sigma^2{\bf I}), \forall k$.
                        \ELSIF{$s>1$}
                            \STATE Set $({\bm \mu}_k^{(0)}, {\bm \Sigma}_k^{(0)}) = ({\bm \mu}^{(I_{EM})}_{k,J^{\star}}, {\bm\Sigma}^{(I_{EM})}_{k,J^{\star}}), \forall k$.
                        \ENDIF
                        \STATE Determine $\hat{p}_{k,j}({\bf y}[n])$ from Steps 5-10 in Algorithm~1. 
                    \ELSIF{KDE-based LF estimation}
                        \IF{$s=1$}
                            \STATE Construct a labeled dataset $\check{\mathcal{D}}_{{\bf y}_j}$ from \eqref{eq:coarse_MLD}.  
                        \ELSIF{$s>1$}
                            \STATE Construct a labeled dataset $\hat{\mathcal{D}}_{{\bf y}_{J^\star}}$ from \eqref{eq:kde_Jstar}. 
                        \ENDIF
                        \STATE Determine $\hat{p}_{k,j}({\bf y}[n])$ from Steps 13-17 in Algorithm~1. 
                    \ENDIF
                \ENDFOR
            
            \STATE Calculate $w_j,\forall j$ from \eqref{eq:aggregation_uniform}, \eqref{eq:aggregation_dirichlet}, or \eqref{eq:aggregation_hard}.
            \STATE Set $\hat{p}_k^\star({\bf y}[n]) = \sum^{J}_{j=1} w_{j} \hat{p}_{k,j}({\bf y}[n]), \forall k, n\in\mathcal{T}_{\rm d,s}$.
            \STATE Determine $\hat{k}^{\star}[n]$ from \eqref{eq:final_ml},  $n\in\mathcal{T}_{\rm d,s}$.
            \STATE Set $\hat{\bf s}[n] = {\bf s}_{\hat{k}^{\star}[n]}, n\in\mathcal{T}_{{\rm d},s}$.
            \ENDFOR
            
        \end{algorithmic}}} 
    \end{algorithm}

    \section{Simulation Results}
    In this section, using simulations, we demonstrate the superiority of the proposed LF estimation method in MIMO systems with hardware impairments. 
        In these simulations, we incorporate least-squares channel estimation using pilot signals. We employ 4-QAM and set $T_{\rm p}=8$, $T_{\rm d}=1000$, and $N_{t}=2$, unless specified otherwise. The SNR of the system is defined as ${\sf SNR} = N_t/\sigma^2$.
        We assume that non-ideal PAs are employed at the transmitter and low-resolution ADCs are employed at the receiver. In this scenario, the TX distortion function captures the effect of the non-ideal PAs and given as in \eqref{eq:PA}. Also, the RX distortion function captures the effect of the low-resolution ADCs and is therefore given as in \eqref{eq:ADC}. To determine these functions, we set $(\alpha_a,\epsilon_a,\alpha_\phi,\epsilon_\phi)=(1.96,0.99,2.53,2.82)$ and  $y_k=-1.75+(k-1)/2$ for $k\in\{1,\cdots,2^{B}\}$ and $b_k=(y_k+y_{k+1})/2$ for $k\in\{1,\cdots,2^B-1\}$ with $B=3$.  
        When considering the time-varying channels, we model the channel matrix using a first-order autoregressive process. Specifically, the channel matrix at time slot $n$ is modeled as
    \begin{align}
        {\mathbf{H}}[n] = \zeta {\mathbf{H}}[n-1] +\sqrt {1-\zeta ^{2}}{\mathbf{V}}[n],
    \end{align}
    for $n \in \{2, \ldots , T\}$, where $\zeta$ is a temporal correlation coefficient, and ${\bf V}[n]$ introduces channel evolution at each time slot, with its entries independently drawn from a complex Gaussian distribution $\mathcal{CN} (0, 1)$. The initial channel state ${\bf H}[1]$ is initialized with entries independently sampled from the same distribution. In our simulations, we set $\zeta=0$ for time-invariant channels and $\zeta = 0.9999$ for time-varying channels. The concentration parameter for probabilistic and max aggregation is set as $\alpha_k=2$ for all $k\in\{1,2,\dots,K\}$.

    For performance comparison, we consider the following methods:
    \begin{itemize}
    \item{\bf Channel-estimation (CE)-based MLD:} This method performs the ML detection based on conventional channel estimation without considering hardware impairments. In particular, this method assumes that $p({\bf y}[n]|{\bf s}_k)= \mathcal{CN}\big({\bf y}[n] ;\hat{\bf H}{\bf s}_k, \sigma^2{\bf I} \big)$, where $\hat{\bf H}_{\rm LS}$ is a channel estimate determined by a least squares method.

    \item{\bf EMNL:} This method was developed in \cite{jinman}, which utilizes all the received signals as noisy training samples and employs a modified EM-based LF estimation method designed for a noisy training dataset. We set $\epsilon = 10^{-8}$ and $I_{\rm EM} = 20$ for the convergence.

    \item{\bf Proposed EM or Proposed KDE:} This is the proposed method summarized in Algorithm 1 (for time-invariant channels) or Algorithm 2 (for time-varying channels). We set $T_{\rm b} = 250$, $I_{\rm DA}=10$, $I_{\rm EM}=10$, $J=9$, $\sigma_G\in\{0.04,0.08,0.12\}$, $\sigma_U\in\{0.8,1,1.2\}$, and $\sigma_L\in\{0.21,0.24,0.27\}$, unless specified otherwise. For time-varying channels, we set the number of sub-blocks as $S=4$. For the weight determination, we employ the probabilistic aggregation method, unless specified otherwise.

    \item{\bf Optimal MLD:} This method performs the ideal ML detection by utilizing the {\em true} LFs. The underlying assumption is that the receiver has perfect knowledge about channels and both TX and RX distortion functions.
    \end{itemize}

    \begin{figure}
        \centering
        {\epsfig{file=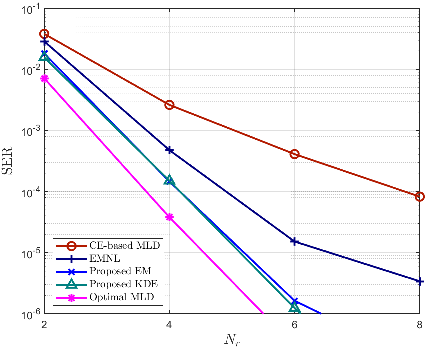, width = 7.5cm}}
        \caption{SER comparison of various detection methods for a different number of receive antennas with ${\sf SNR} = 20$dB in time-invariant channels.} \vspace{-1.5mm}
        \label{fig:inv_nr}
    \end{figure}

    Fig. \ref{fig:inv_nr} compares the symbol error rates (SERs) of various detection methods for a different number of receive antennas with ${\sf SNR} = 20$dB in time-invariant channels. 
    Notably, the MLD method with the proposed LF estimation methods outperforms the existing CE-based MLD and EMNL methods in terms of the SER performance. The performance gain becomes more evident as the number of received antennas increases, particularly in scenarios that serve reliable communication.
    The performance gain of the proposed methods stems from three key factors. Firstly, the data augmentation technique provides a more reliable and diverse training dataset for LF estimation, enabling better generalization to hardware impairment scenarios. Secondly, the boosting process, which calculates a weighted summation of multiple LF estimates, further reduces the LF estimation error. Lastly, an increase in the number of received antennas enables more accurate initialization for the EM algorithm. Improved initialization, combined with our data augmentation and boosting techniques, results in a more robust LF estimation, particularly in scenarios with multiple receive antennas.

    \begin{figure}
        \centering
        {\epsfig{file=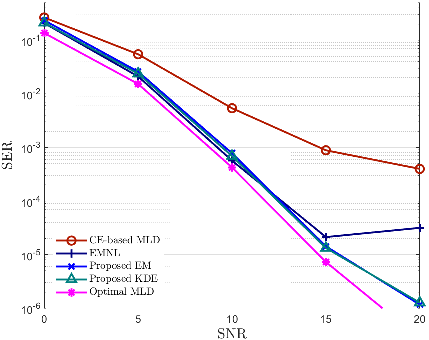, width = 7.5cm}}
        \caption{SER comparison of various detection methods for different SNRs with $(N_t, N_r)=(2,6)$ in time-invariant channels.} \vspace{-1.5mm}
        \label{fig:inv_2by6}
    \end{figure}
    
    Fig. \ref{fig:inv_2by6} compares the SERs of various detection methods for different  SNRs with $(N_t, N_r)=(2,6)$ in time-invariant channels. Fig.~\ref{fig:inv_2by6} shows that the proposed methods achieves significant performance gains, particularly in the high SNR regime. This improvement is especially notable compared to the EMNL approach, which exhibits increasing SER under high SNR conditions. The primary reason for this behavior is that the limited number of data samples leads to an overfitting problem, adversely affecting the performance of the EM algorithm. Our proposed method effectively mitigates this limitation by enhancing the diversity of augmented data samples and leveraging boosting techniques.

    \begin{figure}
        \centering
        {\epsfig{file=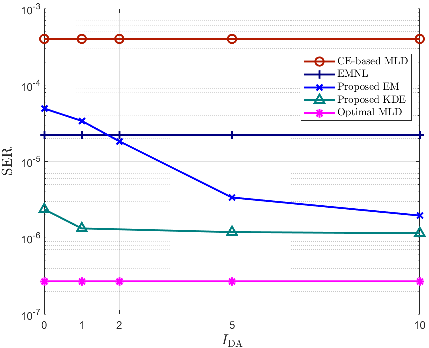, width = 7.5cm}}
        \caption{SER comparison of the proposed LF estimation method with a different number of augmented data samples when $(N_t, N_r)=(2,6)$ and ${\sf SNR} = 20$dB  in time-invariant channels.} \vspace{-1.5mm}
        \label{fig:inv_da}
    \end{figure}
    
    Fig. \ref{fig:inv_da} compares the SER of the proposed LF estimation method with a different number of augmented data samples when $(N_t, N_r)=(2,6)$ and ${\sf SNR} = 20$dB  in time-invariant channels. Fig.~\ref{fig:inv_da} shows that the proposed EM-based and KDE-based methods with a sufficient number of augmented data samples (e.g., $I_{\rm DA}=10$) achieve the closest performance to the optimal MLD method. Without data augmentation (i.e., $I_{\rm DA}=0$), the KDE-based method outperforms the conventional EM-based or CE-based methods because the KDE-based method does not rely on a specific model and therefore can avoid a model mismatch problem. The performance gap between the EM-based and KDE-based methods can be reduced by employing our data augmentation strategy, making the proposed EM-based method an appealing solution for practical systems. All these results demonstrate the effectiveness of our strategy in mitigating both the data shortage and overfitting problems, leading to a better LF estimation performance.
    
    \begin{figure}
        \centering
        {\epsfig{file=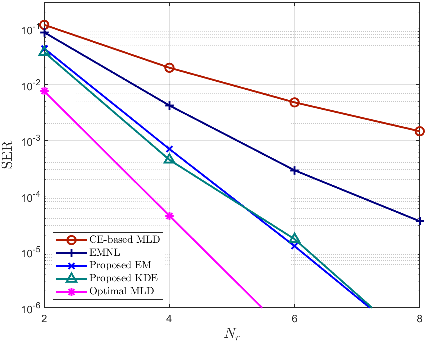, width = 7.5cm}}
        \caption{SER comparison of the proposed method with a different number of receive antennas when SNR is $20$dB in time-varying scenarios.} \vspace{-1.5mm}
        \label{fig:var_nr}
    \end{figure}
    
    Fig. \ref{fig:var_nr} compares the SERs of various detection methods for a different number of receive antennas with ${\sf SNR} = 20$dB in time-varying channels. 
    Compared to the results in Fig.~\ref{fig:inv_nr}, Fig.~\ref{fig:var_nr} shows that the performance gap between the optimal MLD and other conventional methods becomes significantly larger. This is because the mismatches in the LF arise not only from hardware impairments but also from temporal variations within a communication block. Unlike these conventional methods, the proposed methods utilize sub-block-based LF estimation, allowing them to track temporal changes in the LFs under time-varying channel conditions. As a result, the proposed methods effectively mitigate LF mismatches even in time-varying scenarios.

    \begin{figure}
        \centering
        {\epsfig{file=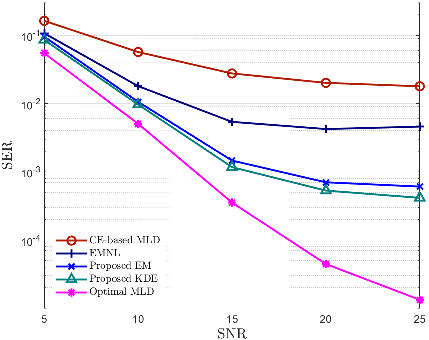, width = 7.5cm}}
        \caption{SER comparison of various detection methods for different SNRs with $(N_t, N_r)=(2,4)$  in time-varying channels.} \vspace{-1.5mm}
        \label{fig:var_2by4}
    \end{figure}
    
    Fig. \ref{fig:var_2by4} compares the SERs of various detection methods for different  SNRs with $(N_t, N_r)=(2,4)$ in time-varying channels. Compared to the results in Fig. \ref{fig:inv_2by6}, Fig. \ref{fig:var_2by4} shows that all detection methods experience significant performance degradation compared to the optimal MLD method, primarily due to the increased mismatch in LFs caused by time-varying channels.
    Even in this challenging scenario, our modified methods described in Sec.~\ref{Sec:Pro_TV} remain effective, demonstrating a significant performance gain over conventional methods. It is also shown that the proposed KDE-based method outperforms the EM-based method, demonstrating significant improvements, particularly in the high-SNR regime. This result suggests that the data-driven approach of the KDE method, as opposed to the model-driven approach of the EM-based method, enables more accurate LF estimation especially when operating on highly reliable received datasets available in high-SNR environments.

    \begin{table}[t]
    \centering
    \caption{SER comparison of the proposed EM-based LF estimation method with various weight determination methods when $(N_t, N_r)=(2,4)$ and ${\sf SNR}=20$dB}\label{table:inv_agg}
    \begin{tabular}{cccc}
    \hline
    \multicolumn{4}{c}{{\bf Time-invariant channels}} \\ \hline
    \multicolumn{1}{c|}{SNR } & Uniform & Probabilistic & Max \\ \hline
    \multicolumn{1}{c|}{10 dB} & ${\bf 8.00\times 10^{-3}}$ &  ${\bf 8.00\times 10^{-3}}$ & $8.65\times 10^{-3}$  \\
    \multicolumn{1}{c|}{15 dB} &  $6.59\times 10^{-4}$ &  ${\bf 6.53\times 10^{-4}}$ &  $7.76\times 10^{-4}$  \\
    \multicolumn{1}{c|}{20 dB} & $1.72\times 10^{-4}$ & ${\bf 1.46\times 10^{-4}}$ & $2.00\times 10^{-4}$  \\
    \multicolumn{1}{c|}{25 dB} & $7.00\times 10^{-5}$ & ${\bf 5.62\times 10^{-5}}$ & $7.51\times 10^{-5}$  \\ \hline
    \multicolumn{4}{c}{{\bf Time-varying channels}} \\ \hline
    \multicolumn{1}{c|}{SNR } & Uniform & Probabilistic & Max \\ \hline
    \multicolumn{1}{c|}{10 dB} & ${\bf 1.06\times 10^{-2}}$ &  ${\bf 1.06\times 10^{-2}}$ & $1.16\times 10^{-2}$ \\
    \multicolumn{1}{c|}{15 dB} &  ${\bf 1.46\times 10^{-3}}$ &  ${\bf 1.46\times 10^{-3}}$ &  $1.73\times 10^{-3}$  \\
    \multicolumn{1}{c|}{20 dB} & $7.03\times 10^{-4}$ & ${\bf 7.01\times 10^{-4}}$ & $8.16\times 10^{-4}$  \\
    \multicolumn{1}{c|}{25 dB} & ${\bf 6.07\times 10^{-4}}$ & ${\bf 6.07\times 10^{-4}}$ & $6.98\times 10^{-4}$  \\ \hline
    \end{tabular}
    \end{table}

    Table~\ref{table:inv_agg} compares the SERs of the proposed EM-based estimation method with various weight determination methods when $(N_t, N_r)=(2,4)$ and ${\sf SNR}=20$dB. The results indicate that the performance of the proposed method remains relatively consistent across all weight determination methods. Nevertheless, the probabilistic aggregation method achieves the best SER performance among three weight determination methods in all cases. This superior performance is attributed to its ability to utilize all LF estimates based on their reliability levels, whereas other methods either disregard reliability or rely on a single estimate. These findings suggest that a well-designed weight determination method can further enhance the performance of the proposed LF estimation method.


    \section{Conclusion}
    In this paper, we have proposed a novel data augmentation and boosting strategy  for LF estimation in MIMO communication systems with hardware impairments. Our strategy effectively addresses data shortages and overfitting issues by generating multiple augmented datasets through noise injection into online received signals and leveraging boosting to combine multiple LF estimates. Furthermore, we have extended our approach to a sub-block-based LF estimation framework, enabling accurate LF estimation even in time-varying channel environments. Simulation results demonstrated the superiority of our strategy compared to existing detection methods in both time-invariant and time-varying MIMO systems with non-ideal PAs and low-resolution ADCs. These findings provide strong evidence that data augmentation combined with boosting is a powerful approach for enhancing data detection performance in the presence of unpredictable and unknown hardware impairments.

    An important direction for future research is to incorporate the proposed data augmentation and boosting strategy with linear data detection methods, in order to support high-order modulation and a large number of transmit antennas. Another important research direction is to extend our strategy to MIMO-OFDM systems with hardware impairments. It would also be possible to improve the performance of our strategy by combining it with advanced learning-based data detection methods, such as deep-learning-based detection methods.
    


    \bibliographystyle{IEEEtran}
    \bibliography{Reference}

\end{document}